\documentclass[preprint,superscriptaddress,amsmath,amssymb,prd,aps,showpacs,floatfix,nofootinbib]{revtex4-1}
\usepackage[T1]{fontenc}
\usepackage{lmodern}

\usepackage[dvipsnames,x11names]{xcolor}

\usepackage[colorlinks=true,linkcolor=Blue,citecolor=ForestGreen,urlcolor=NavyBlue]{hyperref}

\usepackage{graphicx}
\usepackage[sort&compress]{natbib}
\usepackage{lipsum}
\usepackage{morefloats}
\usepackage[pdf]{pstricks}
\usepackage{subfigure}
\usepackage{amsmath}
\usepackage{amssymb}
\usepackage{amsfonts}
\usepackage{xcolor}
\usepackage{rotating}
\usepackage{cancel}
\usepackage{mathtools}
\usepackage{bbm}
\usepackage{dsfont}
\usepackage{bbold}
\usepackage{multirow}
\usepackage{ulem}
\usepackage{physics}
\usepackage{orcidlink}

\newcommand{\GXNU}{Department of Physics, Guangxi Normal University, Guilin 541004, China}
\newcommand{\GXZD}{Guangxi Key Laboratory of Nuclear Physics and Technology, Guangxi Normal University, Guilin 541004, China}
\newcommand{\IFIC}{Departamento de F\'{\i}sica Te\'orica and IFIC, Centro Mixto Universidad de
Valencia-CSIC Institutos de Investigaci\'on de Paterna, Aptdo.22085,
46071 Valencia, Spain}
\newcommand{\CSU}{School of Physics, Central South University, Changsha 410083, China}

\begin{document}

\frenchspacing

\title{\boldmath Correlation functions for the $N^*(1535)$ and the inverse problem}
%\date{\today}

\author{Raquel Molina\orcidlink{0000-0001-9427-240X}}
\email{raquel.molina@ific.uv.es}
\affiliation{\GXNU}%
\affiliation{\IFIC}%

\author{Chu-Wen Xiao\orcidlink{0000-0001-5303-8350}}
\email{xiaochw@gxnu.edu.cn}
\affiliation{\GXNU}%
\affiliation{\GXZD}%
\affiliation{\CSU}%

\author{Wei-Hong Liang\orcidlink{0000-0001-5847-2498}}%
\email{liangwh@gxnu.edu.cn}
\affiliation{\GXNU}%
\affiliation{\GXZD}%

\author{Eulogio Oset\orcidlink{0000-0002-4462-7919}}%
\email{Oset@ific.uv.es}
\affiliation{\GXNU}%
\affiliation{\IFIC}%

\begin{abstract}
The $N^*(1535)$ can be dynamically generated in the chiral unitary approach with the coupled channels, $K^0 \Sigma^+, K^+ \Sigma^0, K^+ \Lambda$ and $\eta p$. In this work we evaluate the correlation functions for every channel and face the inverse problem. Assuming the correlation functions to correspond to real measurements, we conduct a fit to the data within a general framework in order to extract the information contained in these correlation functions. The bootstrap method is used to determine the uncertainties of the different observables, and we find that, assuming errors of the same order than in present measurements of correlation functions, one can determine the scattering length and effective range of all channels with a very good accuracy. Most remarkable is the fact that the method predicts the existence of a bound state of isospin $\frac{1}{2}$ nature around the mass of the $N^*(1535)$ with an accuracy of $6\; \rm MeV$. These results should encourage the actual measurement of these correlation functions (only the $K^+ \Lambda$ one is measured so far), which can shed valuable light on the relationship of the $ N^*(1535)$ state to these coupled channels, a subject of continuous debate.
\end{abstract}

\maketitle

\section{Introduction}
\label{sec:intro}

The $N^*(1535)$ state plays an important role in many processes in hadron physics and is the subject of intense discussion concerning its nature. One of the pioneering works on the chiral unitary approach was devoted to discuss the nature of this state,
which in Refs.~\cite{siegel,weise} was branded as a dynamically generated state from the interaction of coupled channels, $K \Sigma, K \Lambda, \eta N$ and $\pi N$.
A more detailed work with the same conclusions followed in Ref.~\cite{inoue},
with the  $K \Sigma, K \Lambda$ channels playing a major role in the wave function of the state.

From the three quark structure, this state has always been problematic,
since the $N^*(1535)$ is above the first positive parity excitation of the nucleon,
supposed to be the $N^*(1440)$, contrary to simple quark model expectations.
Support for the molecular picture in coupled channels is also found in Refs.~\cite{arriola,liuzou,arriola,bruns,cieply,nacher}.
Other works suggest the mixture of a three quark component with some pentaquark configuration \cite{24,25}.
Analysis of $\phi$ production in proton-proton reactions
suggested a large coupling of the $N^*(1535)$ to $s \bar s$ components \cite{chiangzou,doringzou}
in line with the relevance of the $K \Sigma, K \Lambda$ channels.
The $N^*(1535)$ coupling strength to the $K^+ \Lambda$ channel is extracted
from the $K^+ \Lambda$ photoproduction database in Ref.~\cite{mart}
and found to be consistent with the results of the chiral unitary approach.
The couplings to $K \Lambda$ and $\eta N$ are also obtained in Ref.~\cite{liuzou}
from the study of the $J/\psi$ to $p \bar p \eta$ and $\bar p K^+ \Lambda$,
indicating the relevance of these components in the $N^*(1535)$ state.

Even admitting the importance of the molecular components in the wave function of the $N^*(1535)$ state,
many works suggest that the three quark components are also relevant in the structure of the state \cite{Hyodo:2008xr,Sekihara:2015gvw,sekidos}.
In more recent papers, combining the study in the real space with lattice simulation results \cite{jiajun1,jiajun2},
the conclusion is similar,
but stressing more the weight of the three quark component.
In Ref.~\cite{jiajun1} the $\eta N$ and $\pi N$ channels are considered in addition to the three quark core,
while in Ref.~\cite{jiajun2} the $K \Lambda$ component is included in addition.
The $K \Sigma$ channel that plays an essential role in the chiral unitary approach is not included in these two latter works.
Another issue related to the $N^*(1535)$ is the problem of eta-nucleus interaction and the study of possible eta bound states in nuclei,
which has also attracted much attention \cite{metag}.

The purpose of the present work is to bring the attention to a new source of information so far not exploited.
This is the information that can be obtained about the relevance of the meson-baryon components of the $N^*(1535)$
from the study of the interaction of these channels themselves.
Obviously we cannot implement an experiment making $K \Lambda$ or $K \Sigma$ scattering,
but this information can nowadays be obtained via the study of the correlation functions of these channels.
In this case, in $p p$, $p A$ and $A A$ collisions,
pairs of the particles that we are interested in are measured at low relative momenta
and from there one can obtain the observables that would be determined in collisions of these pairs.
Performing an analytical extrapolation of the results below threshold
one can determine if a bound state appears related to the interaction of the pairs studied.
Nothing can be more instructive about the role the $K \Sigma$ and $K \Lambda$ components in the $N^*(1535)$ build up,
than showing that from the knowledge of the $K \Sigma$ and $K \Lambda$ interaction in a moderate range of energies above their respective thresholds, the existence of the $N^*(1535)$ state and its properties can be determined with a remarkable precision.

Correlation functions for the channels needed here are not available yet,
but very close information on $\bar K N$ interaction and related channels is already known
\cite{STAR:2014dcy,ALICE:2017jto,STAR:2018uho,ALICE:2018ysd,ALICE:2019hdt,ALICE:2019eol,
ALICE:2019buq,ALICE:2019gcn,ALICE:2020mfd,ALICE:2021szj,ALICE:2021cpv,Fabbietti:2020bfg}.
Theoretical work on the issue is also already available \cite{Morita:2014kza,Ohnishi:2016elb,Morita:2016auo,Hatsuda:2017uxk,Mihaylov:2018rva,Haidenbauer:2018jvl,Morita:2019rph,
Kamiya:2019uiw,Kamiya:2021hdb,Liu:2023uly,Albaladejo:2023pzq,Ikeno:2023ojl,Torres-Rincon:2023qll,Liu:2023wfo}.

While most of the theoretical work compares results of models with existing correlation functions,
it has only been recently that the inverse problem has been given a solution.
This means that starting from experimental correlation functions,
a method is proposed that allows to determine the observables of the interaction,
like scattering lengths and effective ranges and the eventual existence of a bound state of the components studied \cite{Ikeno:2023ojl,tccinv,tbbinv}.
We follow this line of work. In the absence of the needed correlation functions,
we construct them from the successful chiral unitary approach
and take synthetic data from these correlation functions making fits to these data to determine the parameters of the inverse method,
from where we evaluate the observables of the systems.
While, certainly,  the method should return an output consistent with the input used to determine the correlation functions,
the important information that we obtain,
for which we use the resampling or bootstrap method,
is the uncertainty by means of which we can determine these observables using only the correlation functions as input.
In the present case, we shall evaluate correlation functions for $K \Sigma, K \Lambda$ and $\eta N$
and we shall see that the scattering lengths and effective ranges can be determined with high accuracy from them. Furthermore, we will conclude the existence of the $N^*(1535)$ as a bound state of the  $K \Sigma, K \Lambda$  components with a precision of a few MeV.
This is quite remarkable, being the  $N^*(1535)$ lying about $75 \; \rm MeV$ below the $K \Lambda$
threshold and $150 \; \rm MeV$ below the $K \Sigma$ one.
This indicates the high potential existing in the correlation functions,
which should encourage experimental work in this direction to bring light into the permanent debate about the nature of many hadronic states.

\section{Formalism}
\subsection{The chiral unitary approach for the $N^*(1535)$}
In Ref.~\cite{inoue}, the coupled channels $K \Sigma, K \Lambda, \pi N, \eta N$ were considered and the zero charge states were taken,
namely,
\begin{equation}\label{eq:channels1}
	K^+\Sigma^-,\; K^0 \Sigma^0,\; K^0 \Lambda, \;\pi^- p, \;\pi^0 n, \;\eta n.
\end{equation}
This choice corresponds to the states with isospin $| I,I_3=-\frac{1}{2}\rangle$ with $I=\frac{1}{2}, \frac{3}{2}$.
These states are not convenient for correlation functions since one has the Coulomb interaction
which makes the calculations more involved
and the experimental correlation functions at low momenta are dominated by the Coulomb interaction.
Then, we take those corresponding to $| I,I_3=\frac{1}{2}\rangle$,
\begin{equation}\label{eq:channels2}
	K^0\Sigma^+,\; K^+ \Sigma^0,\; K^+ \Lambda, \;\pi^+ n, \;\pi^0 p, \;\eta p.
\end{equation}
In addition, since we are only interested about energies which are above the thresholds of $K^0\Sigma^+$, $K^+ \Sigma^0$, $K^+ \Lambda$ and $\eta p$,
we can safely neglect the $\pi N$ channels which are about $400-600 \; \rm MeV$ below in energy.
The $\pi N$ channel is also about $450\, \rm MeV$ below the nominal mass of the $N^*(1535)$, and with such differences in mass,
these far away channels have a negligible role in the mass of the dynamically generated states,
affecting only their width.
In this sense, with the purpose of calculating the position of the $N^*(1535)$ resonance from the knowledge of the correlation functions,
we can simplify the approach eliminating the $\pi^+ n$ and $\pi^0 p$ channels,
knowing that we will pay a small price in the determination of the width of the $N^*(1535)$, and also the imaginary parts of the scattering length and effective range of the $\eta p$ channel.
However, here these are secondary issues in our aim,
which is to establish the link between the $N^*(1535)$ and the $K\Sigma, K\Lambda$ components.

Following the steps of the chiral unitary approach of Ref.~\cite{inoue},
from the chiral Lagrangians one obtains an interaction between the channels of Eq.~\eqref{eq:channels2},
\begin{equation}\label{eq:Vij}
	V_{ij}=-\dfrac{1}{4f^2} C_{ij} (k^0+k^{\prime \,0});~~~~ f=93\, \rm MeV,
\end{equation}
with $k^0, k^{\prime \,0}$ the energies of the initial and final mesons,
\begin{equation}\label{eq:k0}
k^0 =\dfrac{s+m_1^2-M_1^2}{2\sqrt{s}};~~~~ k^{\prime \,0} =\dfrac{s+m_2^2-M_2^2}{2\sqrt{s}},
\end{equation}
where $m_1, M_1$ are the masses of the initial meson, baryon and $m_2, M_2$ the same for the final ones.
The matrix $C_{ij}$ is given in Table \ref{tab:Cij}.
\begin{table}[t]
	\caption{$C_{ij}$ coefficients of Eq.~\eqref{eq:Vij}.}
\centering
\begin{tabular*}{0.75\textwidth}{@{\extracolsep{\fill}}c| c c c c c c}
\toprule
$C_{ij}$         & $K^0 \Sigma^+$  &  $K^+ \Sigma^0$  &  $K^+ \Lambda$  &  $\pi^+ n$  &  $\pi^0 p$               &  $\eta p$  \\
\hline
$K^0 \Sigma^+$   & $1$             &  $\sqrt{2}$      &  $0$            &  $0$        &  $\frac{1}{\sqrt{2}}$   &  $-\sqrt{\frac{3}{2}}$ \\[2mm]
$K^+ \Sigma^0$   &       & $0$   &  $0$ &  $\frac{1}{\sqrt{2}}$    &  $-\frac{1}{2}$  &   $-\frac{\sqrt{3}}{2}$ \\[2mm]
$K^+ \Lambda$    &   &  &  $0$  &  $-\sqrt{\frac{3}{2}}$  &  $-\frac{\sqrt{3}}{2}$  &  $-\frac{3}{2}$ \\[2mm]
$\pi^+ n$          &   &   &   & $1$  & $\sqrt{2}$   &  $0$  \\
$\pi^0 p$        &   &   &   &   & $0$  & $0$ \\
$\eta p$         &   &   &   &   &   & $0$ \\
\hline\hline
\end{tabular*}
\label{tab:Cij}
\end{table}

We should not overlook one important detail.
The $K^+\Lambda$ has no diagonal interaction and does not couple to $K^0 \Sigma^+$ or $K^+ \Sigma^0$.
Its role in the problem is through its coupling to the $\eta p$ channel.
Note that only $K^0\Sigma^+$ has a diagonal attractive interaction.
Hence this is the essential channel that can create a bound state,
and one finds that omitting the $K\Sigma$ channels no pole is obtained in the approach.

We include all channels in Table \ref{tab:Cij}, although we shall omit the $\pi N$ channels in the calculations.
With the information of Eq.~\eqref{eq:Vij}, we use the Bethe-Salpeter (BS) equation in coupled channels
\begin{equation}\label{eq:BS}
  T=[1-VG]^{-1} \, V,
\end{equation}
with $G$ the diagonal loop function ${\rm diag}(G_i)$ and $G_i$ regularized with a cutoff in three momentum given by
\begin{equation}\label{eq:G}
  G_i(s)= \int_{|\vec q\,| < q_{\rm max}} \dfrac{{\rm d}^3 q}{(2\pi)^3} \; \dfrac{\omega_i(q)+E_i(q)}{2\,\omega_i(q)\, E_i(q)}\; \dfrac{2M_i}{s-[\omega_i(q)+E_i(q)]^2 + i \varepsilon},
\end{equation}
with $\omega_i(q)=\sqrt{\vec q^{\;2} +m_i^2}$, $E_i(q)=\sqrt{\vec q^{\;2} +M_i^2}$, and $m_i, M_i$ the mass of the meson, baryon in the loop.

We use a value of $q_{\rm max}= 630 \, \rm MeV$,
the same one used in Ref.~\cite{osetramos} to study the $\bar K N$ interaction with its coupled channels,
which gives rise to the two $\Lambda(1405)$ states.
The $T$ matrix of Eq.~\eqref{eq:BS} shows a pole at $1515\, \rm MeV$,
close to the pole position of the $N^*(1535)$ in the PDG \cite{pdg} ($1510\, \rm MeV$).

For later use in the paper we write here the isospin wave functions with the phase convention for the isospin multiplets given as
$(K^+, K^0)$, $(-\pi^+, \pi^0, \pi^-)$, $(-\Sigma^+, \Sigma^0, \Sigma^-)$ and $(p, n)$,
\begin{equation}\label{eq:iso}
  \begin{split}
   |K\Sigma, I=\frac{1}{2}, I_3=\frac{1}{2} \rangle =& \sqrt{\frac{2}{3}} K^0 \Sigma^+ + \sqrt{\frac{1}{3}} K^+ \Sigma^0, \\[1.5mm]
   |K\Sigma, I=\frac{3}{2}, I_3=\frac{1}{2} \rangle =& -\sqrt{\frac{1}{3}} K^0 \Sigma^+ + \sqrt{\frac{2}{3}} K^+ \Sigma^0.
  \end{split}
\end{equation}

\subsection{Correlation functions}
We follow here the formalism developed in Ref.~\cite{29}, which leads to a slightly modified version of the Koonin-Pratt formula \cite{Koonin,Pratt,Bauer},
takes into account explicitly the range of the interaction in momentum space and is technically easier to implement.
Taking unity for the weights of the different channels, as demanded for an elastic channel, the formulas that we obtain are
\begin{eqnarray}\label{eq:C1}
  C_{K^0\Sigma^+} (p_{K^0})&=& 1+4\,\pi\, \theta(q_{\rm max}-p_{K^0})\, \int dr \, r^2 S_{12}(r)  \cdot \nonumber\\[2mm]
  && \left\{ \left|j_0(p_{K^0}\, r)+T_{K^0 \Sigma^+, K^0 \Sigma^+}(E)\; \tilde{G}^{(K^0\Sigma^+)}(r; E)\right|^2 \right.  \nonumber\\[2mm]
  &&\;+ \left|T_{K^+ \Sigma^0, K^0 \Sigma^+}(E)\; \tilde{G}^{(K^+\Sigma^0)}(r; E) \right|^2
  +  \left|T_{K^+ \Lambda, K^0 \Sigma^+}(E)\; \tilde{G}^{(K^+\Lambda)}(r; E) \right|^2     \nonumber\\[2mm]
  &&\;\left. +\left|T_{\eta p,  K^0 \Sigma^+}(E)\; \tilde{G}^{(\eta p)}(r; E)\right|^2 - j_0^2 (p_{K^0}\, r)\right\},  %\nonumber\\[2mm],
\end{eqnarray}
\begin{eqnarray}\label{eq:C2}
  C_{K^+\Sigma^0} (p_{K^+})&=& 1+4\,\pi\, \theta(q_{\rm max}-p_{K^+})\, \int dr \, r^2 S_{12}(r)  \cdot \nonumber\\[2mm]
  && \left\{ \left|j_0(p_{K^+}\, r)+T_{K^+ \Sigma^0, K^+ \Sigma^0}(E)\; \tilde{G}^{(K^+\Sigma^0)}(r; E)\right|^2 \right.  \nonumber\\[2mm]
  &&\;+ \left|T_{K^0 \Sigma^+, K^+ \Sigma^0}(E)\; \tilde{G}^{(K^0\Sigma^+)}(r; E) \right|^2
  +  \left|T_{K^+ \Lambda, K^+ \Sigma^0}(E)\; \tilde{G}^{(K^+\Lambda)}(r; E) \right|^2     \nonumber\\[2mm]
  &&\;\left. +\left|T_{\eta p, K^+ \Sigma^0}(E)\; \tilde{G}^{(\eta p)}(r; E)\right|^2 - j_0^2 (p_{K^+}\, r)\right\},  %\nonumber\\[2mm],
\end{eqnarray}
\begin{eqnarray}\label{eq:C3}
  C_{K^+\Lambda} (p_{K^+})&=& 1+4\,\pi\, \theta(q_{\rm max}-p_{K^+})\, \int dr \, r^2 S_{12}(r)  \cdot \nonumber\\[2mm]
  && \left\{ \left|j_0(p_{K^+}\, r)+T_{K^+ \Lambda, K^+ \Lambda}(E)\; \tilde{G}^{(K^+\Lambda)}(r; E)\right|^2 \right.  \nonumber\\[2mm]
  &&\;+ \left|T_{K^0 \Sigma^+, K^+ \Lambda}(E)\; \tilde{G}^{(K^0\Sigma^+)}(r; E) \right|^2
  +  \left|T_{K^+ \Sigma^0, K^+ \Lambda}(E)\; \tilde{G}^{(K^+\Sigma^0)}(r; E) \right|^2     \nonumber\\[2mm]
  &&\;\left. +\left|T_{\eta p, K^+ \Lambda}(E)\; \tilde{G}^{(\eta p)}(r; E)\right|^2 - j_0^2 (p_{K^+}\, r)\right\},  %\nonumber\\[2mm],
\end{eqnarray}
\begin{eqnarray}\label{eq:C4}
  C_{\eta p} (p_{\eta})&=& 1+4\,\pi\, \theta(q_{\rm max}-p_{\eta})\, \int dr \, r^2 S_{12}(r)  \cdot \nonumber\\[2mm]
  && \left\{ \left|j_0(p_{\eta}\, r)+T_{\eta p, \eta p}(E)\; \tilde{G}^{(\eta p)}(r; E)\right|^2 \right.  \nonumber\\[2mm]
  &&\;+ \left|T_{K^0 \Sigma^+, \eta p}(E)\; \tilde{G}^{(K^0\Sigma^+)}(r; E) \right|^2
  +  \left|T_{K^+ \Sigma^0, \eta p}(E)\; \tilde{G}^{(K^+\Sigma^0)}(r; E) \right|^2     \nonumber\\[2mm]
  &&\;\left. +\left|T_{K^+ \Lambda, \eta p}(E)\; \tilde{G}^{(K^+ \Lambda)}(r; E)\right|^2 - j_0^2 (p_{\eta}\, r)\right\},  %\nonumber\\[2mm],
\end{eqnarray}
where $p_i$ is the momentum of the particles in the rest frame of the pair,
\begin{equation}\label{eq:pi}
  p_i=\dfrac{\lambda^{1/2}(s, m_i^2, M_i^2)}{2\, \sqrt{s}},
\end{equation}
with $m_i, M_i$ the masses of the meson, baryon of the channel considered.

In the former formulas the $\tilde{G}^{(i)}(r; E)$ function is defined as
\begin{equation}\label{eq:G2}
  \tilde{G}^{(i)}(r; E)= \int \dfrac{{\rm d}^3 q}{(2\pi)^3} \,2M_i \; \dfrac{\omega_i(q)+E_i(q)}{2\,\omega_i(q)\, E_i(q)}\; \dfrac{j_0(q\, r)}{s-[\omega_i(q)+E_i(q)]^2 + i \varepsilon},
\end{equation}
where $\omega_i(q), E_i(q)$ are as defined below Eq.~\eqref{eq:G} and $E=\sqrt{s}$.
Note that $\tilde{G}^{(i)}(r; E)$ corresponds to the loop function $G_i(s)$ of Eq.~\eqref{eq:G} with an extra factor $j_0(q\, r)$ in the numerator.
The function $S_{12}(r)$ is the source function,
accounting for the probability that the pair investigated is produced at a relative distance $r$.
It is commonly parameterized as a Gaussian normalized to $1$,
\begin{equation}\label{eq:S12}
  S_{12}(r)= \dfrac{1}{(\sqrt{4\pi}\, R)^3}  \; e^{-(r^2/4R^2)},
\end{equation}
and $R$ provides the range of the extension of this source.
Typical values of $R$ are $1 \, \rm fm$ for $pp$ collisions, $2-3 \, \rm fm$ for $pA$ collisions and $5\, \rm fm$ for $AA$ collisions.
We shall do the calculations with $R=1\, {\rm fm}$.

There is a convenient trick to save computing time, most welcome in the fits of the bootstrap that we do,
which is to write $\tilde{G}(r; E)$ as
\begin{eqnarray}\label{eq:G3}
  \tilde{G}(r; E)&=& \int \dfrac{{\rm d}^3 q}{(2\pi)^3} \,2M \; \dfrac{\omega(q)+E(q)}{2\,\omega(q)\, E(q)}\; \dfrac{j_0(q\, r)-j_0(q_{\rm on}\, r)}{s-[\omega(q)+E(q)]^2 + i \varepsilon} \nonumber\\[2mm]
  && + j_0(q_{\rm on}\, r)\int \dfrac{{\rm d}^3 q}{(2\pi)^3} \,2M \; \dfrac{\omega(q)+E(q)}{2\,\omega(q)\, E(q)}\; \dfrac{1}{s-[\omega(q)+E(q)]^2 + i \varepsilon},
\end{eqnarray}
where $q_{\rm on}$ is the on-shell value of the momentum of the pair, Eq.~\eqref{eq:pi}.
The reason for this is that the first term is no longer singular
since both the numerator and denominator go to zero as $q \to q_{\rm on}$ and the limit is finite.
Then the second term, that requires precision because of the zero in the denominator when $\varepsilon \to 0$,
can be calculated analytically using the formula of Ref.~\cite{pelaez} (second erratum).

\begin{figure}[b]
\begin{center}
\includegraphics[scale=0.85]{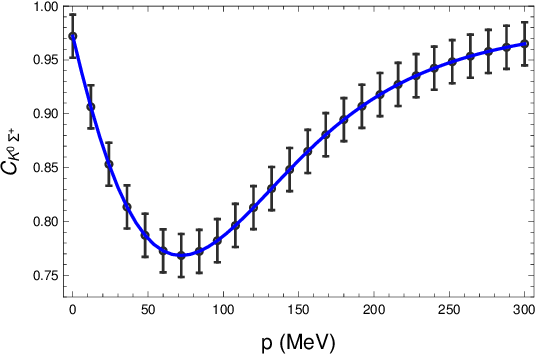}~~~~
\includegraphics[scale=0.85]{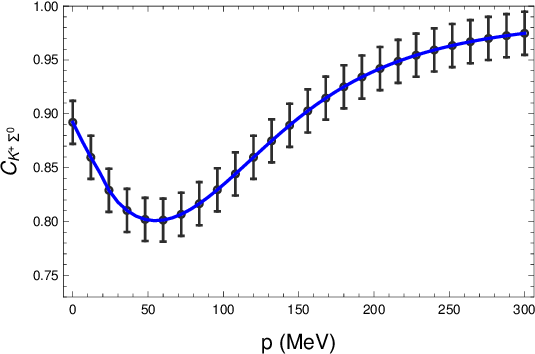}
\end{center}
\vspace{-0.7cm}
\caption{Correlation functions for $K^0\Sigma^+$ and $K^+\Sigma^0$ channels.}
\label{Fig:fig1}
\end{figure}

\begin{figure}[t]
\begin{center}
\includegraphics[scale=0.85]{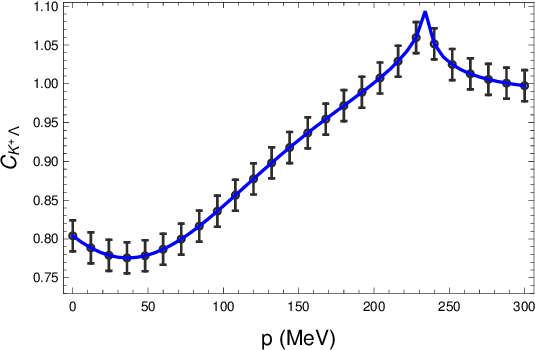}~~~~
\includegraphics[scale=0.85]{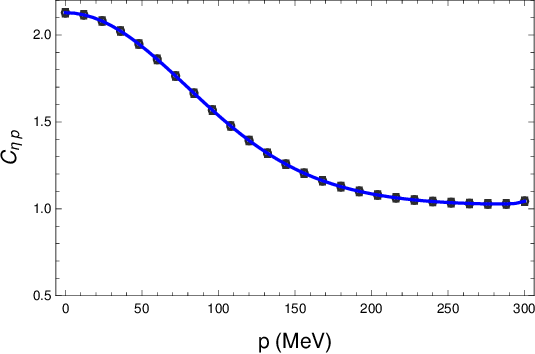}
\end{center}
\vspace{-0.7cm}
\caption{Correlation functions for $K^+\Lambda$ and $\eta p$ channels.}
\label{Fig:fig2}
\end{figure}

With the $T$ matrices obtained with Eqs.~\eqref{eq:Vij},\eqref{eq:BS},
we can construct the correlation functions for the $K^0\Sigma^+, K^+\Sigma^0, K^+\Lambda$ and $\eta p$.
We show the results in Figs.~\ref{Fig:fig1} and \ref{Fig:fig2} calculated with $R=1\, \rm fm$.
As we can see,
the correlation functions for $K^0\Sigma^+$ and $K^+\Sigma^0$ are very similar,
yet not identical.
The reason is dynamical because the weight in the relevant $I=\frac{1}{2}$ component of each pair is different (see Eq.~\eqref{eq:iso}).
The $K^+\Lambda$ correlation function is also different and shows a cusp when the $K\Sigma$ channel opens.
Finally, the $\eta p$ correlation function is very different to the other ones.
It is interesting to see which information is contained in these correlation functions,
and we address this issue in the next subsection.

\subsection{Inverse problem}
We assume now that the correlation functions of Figs.~\ref{Fig:fig1} and \ref{Fig:fig2} are the experimental ones
and we produce some synthetic data by taking $26$ points over the curves with an error of $\pm 0.02$,
typical of the measurements in present correlation functions.
At present there are only data on the $K^+ \Lambda$ correlation function \cite{expelamb} and the errors are even smaller than those assumed here, however, there are errors in the values of $p$.
We assume these bigger errors for the unmeasured correlation functions to be on the safe side in the predictions.
Then, we carry a fit to the data to determine the parameters of the formalism that we discuss below
and with these parameters we determine the different observables.
To get the synthetic points close to a real experiment,
we apply the parametric bootstrap method \cite{bootstrap} generating random centroids with a Gaussian distribution generator,
performing new fits to these data and evaluating observables with the new parameters.
We repeat the procedure about $50$ times and then calculate the averages and dispersion of the different observables.
The reason to do the bootstrap is because when there are correlations between the parameters, the MINUIT errors of these parameters are large, reflecting the presense of these correlations. Using these uncertainties to evaluate uncertainties in the observables is, then, not proper.
The bootstrap method provides the actual uncertainties in the observables. 
The errors obtained with this method are most relevant, since they tell us
with which precision we can expect to know the different observables once the actual correlations are used.
The method also has the advantage that one deals well with the existing correlations between the parameters.
The values of the parameters in each fit can be different,
because of these correlations, but what matters is the precision of the observables obtained.

For the inverse problem we assume that there is a general interaction between the channels
and we implement in it the energy dependence of the chiral unitary approach of Eq.~\eqref{eq:Vij}.
This step is not necessary but will allow us to compare the results of the inverse method with the input used.
Hence, we assume an interaction like
\begin{equation}\label{eq:Vij2}
	V_{ij}=-\dfrac{1}{4f^2} \tilde{C}_{ij} (k^0+k^{\prime \,0}),
\end{equation}
with $\tilde{C}_{ij}$ symmetrical and free parameters.
Then, we would have $10$ parameters plus $q_{\rm max}$ and $R$ to fit the four correlation functions.
Yet, we will impose isospin symmetry for $V_{ij}$ (it is slightly broken in the $T$ matrix due to different masses of the physical particle of the same isospin multiplet).
This means that
\begin{equation}\label{eq:VKSigma}
  \begin{split}
   &\langle K\Sigma, I=\frac{3}{2}, I_3=\frac{1}{2}\big| V  \big| K\Sigma, I=\frac{1}{2}, I_3=\frac{1}{2} \rangle =0 , \\[2mm]
   &\langle K\Sigma, I=\frac{3}{2}, I_3=\frac{1}{2}\big| V  \big|K^+\Lambda \rangle=0, \\[2mm]
   &\langle K\Sigma, I=\frac{3}{2}, I_3=\frac{1}{2}\big| V  \big| \eta p \big\rangle=0.
  \end{split}
\end{equation}
Using these constraints, our $V_{ij}$ matrix takes the form
\begin{equation}\label{eq:Vij4}
  V_{ij}= \left(
    \begin{array}{cccc}
      V_{11} & \sqrt{2}(V_{11}-V_{22}) & V_{13} & V_{14} \\[2mm]
    & V_{22}& \frac{1}{\sqrt{2}}V_{13}  & \frac{1}{\sqrt{2}}V_{14} \\[2mm]
     & & V_{33}  & V_{34} \\[2mm]
     & &   & V_{44} \\
    \end{array}
      \right),
\end{equation}
with the channels $K^0\Sigma^+, K^+\Sigma^0, K^+\Lambda, \eta p$ labeling as $1, 2, 3, 4$.
Our free parameters are now, $C_{11}, C_{22}, C_{13}, C_{14}, C_{33}, C_{34}, C_{44}$ plus $q_{\rm max}$ and $R$.
In total we have $9$ parameters, which can be determined from the four correlation functions.
With the potential of Eq.~\eqref{eq:Vij4}, we construct the $T$ matrices via Eq.~\eqref{eq:BS} and with these matrices the correlation functions.
A fit to the pseudodata returns the values of the free parameters and with them we evaluate the observables as described in the new subsection.

\subsection{Determination of $a, r_0$, binding of state and couplings}
The $T$ matrix of Eq.~\eqref{eq:BS} has a different normalization than the standard one used in Quantum Mechanics.
We have
\begin{equation}\label{eq:Tqm}
T=-\dfrac{8\pi \,\sqrt{s}}{2\, M}\; f^{\rm QM} \simeq -\dfrac{8\pi \,\sqrt{s}}{2\, M}\; \dfrac{1}{-\frac{1}{a}+\frac{1}{2}r_0 k^2 -ik},
\end{equation}
where in the last step we have used the effective range expansion.
Since ${\rm Im} \, T^{-1} = {\rm Im} \, (V^{-1}-G) = -{\rm Im} \,G =\frac{2M}{8\pi \sqrt{s}} k$
(we use one channel for simplicity in this derivation), we see that the term $-ik$ in Eq.~\eqref{eq:Tqm} is reproduced by this conversion factor.

Then we obtain for each channel $j$
\begin{equation}\label{eq:aj}
-\dfrac{1}{a_j}+ \frac{1}{2} \, r_{0, j}\, k_j^2 \equiv -\dfrac{8\pi \,\sqrt{s}}{2\, M_j}\; (T_{jj})^{-1}+ik_j,
\end{equation}
from where we obtain
\begin{equation}\label{eq:aj2}
-\dfrac{1}{a_j}=\left. -\dfrac{8\pi \,\sqrt{s}}{2\, M_j}\; (T_{jj})^{-1}\right|_{\sqrt{s_{{\rm th}, j}}},
\end{equation}
\begin{eqnarray}\label{eq:r0j}
r_{0, j}&=&2 \dfrac{\partial}{\partial k^2}\,
\left[ -\dfrac{8\pi \,\sqrt{s}}{2\, M_j}\; (T_{jj})^{-1}+ik_j \right]_{\sqrt{s_{{\rm th}, j}}},\nonumber\\[2mm]
&=& \dfrac{1}{\mu_j}\; \dfrac{\partial}{\partial \sqrt{s}}\,
\left[ -\dfrac{8\pi \,\sqrt{s}}{2\, M_j}\; (T_{jj})^{-1}+ik_j \right]_{\sqrt{s_{{\rm th}, j}}},
\end{eqnarray}
with $\sqrt{s_{{\rm th}, j}}$ the threshold energy for the channel $j$,
and $\mu_j$ the reduced mass of the two particles of channel $j$.

The pole for a possible bound state is looked for in the second Riemann sheet which is obtained by using $G^{(\rm II)}(\sqrt{s})$ as
\begin{equation}\label{eq:GII}
G^{(\rm II)}(\sqrt{s})= G(\sqrt{s}) + i \dfrac{2M}{4\pi \sqrt{s}}\; q_{\rm on},~~~~q_{\rm on}=\dfrac{\lambda^{1/2}(s, m^2, M^2)}{2\sqrt{s}}
\end{equation}
for channels where ${\rm Re}\, \sqrt{s} > \sqrt{s_{{\rm th}}}$.

Once the pole is determined, the width is approximately two times the imaginary part of the pole (its absolute value),
and the couplings are obtained as
\begin{equation}\label{eq:gj}
g_j^2=\lim_{\sqrt{s} \to \sqrt{s_p}} (\sqrt{s}-\sqrt{s_p})\;T_{jj},
\end{equation}
\begin{equation}\label{eq:gigj}
g_i g_j=\lim_{\sqrt{s} \to \sqrt{s_p}} (\sqrt{s}-\sqrt{s_p})\;T_{ij},
\end{equation}
with $\sqrt{s_p}$ the value of the pole position.
The latter equation allows us to determine relative signs of the couplings for different channels,
and one of them is chosen arbitrarily.

With the value of the couplings, one evaluates the probabilities and wave function at the origin \cite{danijuan,hyodo},
\begin{equation}\label{eq:Pi}
\mathcal{P}_i= g_i^2 \; \dfrac{\partial G_i}{\partial E}; ~~~~~~ \psi(r=0)=g_i\, G_i.
\end{equation}
This is strict for a bound state,
but there are caveats in the case of open channels to which we will come back below.

\section{Results}
As we mentioned above, there are correlations between the parameters,
in particular $q_{\rm max}$ and the strength of the potential.
Since in each fit of the bootstrap method we obtain the value of the parameters,
we show in Table \ref{tab:Cijfit} the average values and errors obtained for $C_{ij}$, $q_{\rm max}$ and $R$.
\begin{table}[h]
	\caption{Values obtained for parameters $C_{ij}$, $q_{\rm max}$ and $R$. The channels are $K^0 \Sigma^+ (1)$, $K^+\Sigma^0$, $K^+ \Lambda (3)$, $\eta p (4)$.}
\centering
\begin{tabular*}{0.75\textwidth}{@{\extracolsep{\fill}}|c| c| c| c| c|}
\toprule
$C_{11}$         & $C_{22}$  &  $C_{33}$  &  $C_{44}$  &  $C_{13}$  \\
$1.10\pm 0.20$   & $-0.02\pm 0.20$      & $0.14\pm 0.30$   &  $0.16\pm 0.07$ &  $0.13\pm 0.20$  \\[2mm]
\hline\hline
 $C_{14}$  & $C_{34}$  & $q_{\rm max} \;(\rm MeV)$  &  $R \;(\rm fm)$  & \\
  $-1.10\pm 0.20$  &   $-1.37\pm 0.16$ & $637\pm 72$  & $1.02\pm 0.02$  & \\
\hline\hline
\end{tabular*}
\label{tab:Cijfit}
\end{table}

The first comment is that one can evaluate $R$ with high precision,
of the order of $2\%$.
The second is that $q_{\rm max}$ is obtained with values around the original one of $630\, \rm MeV$
with uncertainty of about $10\%$.
The matrix elements which are zero in Table \ref{tab:Cij} are small here,
and basically compatible with zero within errors.
$C_{11}$ is of the order of $1$ as in Table \ref{tab:Cij},
$V_{14}$ and $V_{34}$ are both negative and compatible with Table \ref{tab:Cij} within errors
and $V_{34}/V_{14}$ is also compatible with $\sqrt{\frac{3}{2}}$ as in the table.
The errors in the dominant terms are of the order of $20\%$.
This does not mean that we will have errors of this order of magnitude in the observables.
It reflects the correlations between the parameters.
For instance, in one channel we would have $T^{-1}=V^{-1}-G$ and we can make changes in $V^{-1}$ and $G$ (through $q_{\rm max}$) simultaneously,
such that $V^{-1}-G$ does not change at the pole for instance,
and obtain similar results.

The relevant results are those for the observables.
In Table \ref{tab:ai} we show the results of the scattering lengths for the different channels.
\begin{table}[t]
	\caption{Scattering lengths for channel $i$. [in units of fm]}
\centering
\begin{tabular*}{0.7\textwidth}{@{\extracolsep{\fill}}c| c }
\toprule
$a_{1}$         & $a_{2}$     \\
%\hline
~~~$(0.46\pm 0.04) -(0.64\pm 0.03) \, i$~~   & ~~$(0.32\pm0.01) -(0.35\pm 0.02) \, i$ ~~     \\%[2mm]
\hline\hline
  $a_{3}$  &  $a_{4}$ \\
%  \hline
   ~~~  $(0.30\pm0.02) -(0.22\pm 0.04) \,i$~~~   &  $(-0.780\pm 0.013)+ (0\pm 0) \,i$ \\
\hline\hline
\end{tabular*}
\label{tab:ai}
\end{table}
We can see that the errors are smaller than $10\%$ in most cases.

In Table \ref{tab:ri} we see the values of the effective ranges which are determined with a little bigger uncertainty but are still significative.
\begin{table}[b]
	\caption{Effective range parameters for channel $i$. [in units of fm]}
\centering
\begin{tabular*}{0.99\textwidth}{@{\extracolsep{\fill}}c| c| c| c }
\toprule
$r_1$         & $r_2$  &  $r_3$  &  $r_4$  \\
$(-1.1\pm 0.2) -(2.7\pm 0.2) \, i$   & $(-6.2\pm1.4)+ (8.8\pm 0.5) \, i$      & $(-2.8\pm 0.3) -(0.3\pm 0.6) \, i$   &  $-1.48\pm 0.13$  \\
\hline\hline
\end{tabular*}
\label{tab:ri}
\end{table}

Finally, in Table \ref{tab:gi4}
\begin{table}[t]
     \renewcommand{\arraystretch}{0.9}
     \setlength{\tabcolsep}{0.3cm}
\centering
\caption{ Pole position and couplings. [in units of MeV]}
\label{tab:gi4}
\begin{tabular}{c|c|c}
\hline
\hline
$\sqrt{s_p}$ & $g_1$  & $g_2$  \\
   $(1515\pm 6) -(89\pm 9) \,i$  & $(3.7\pm0.3) -(1.04\pm 0.13) \, i$ & $(2.6\pm 0.2) -(0.74\pm 0.10) \, i$   \\
   \cline{2-3}
             & $g_3$ & $g_4$     \\
					   & $(3.6\pm 0.2) -(0.28\pm 0.05) \, i$ & $(-2.68\pm 0.13)+ (1.4\pm 0.2) \,i$    \\
\hline\hline
\end{tabular}
\end{table}
we show the value of the energy for the bound state that we obtain and the couplings of that state to each channel.
It is remarkable to see that the correlation functions contain enough information to deduce
that there is a bound state in the $K\Sigma, K\Lambda$ channels with an energy around $1515 \, \rm MeV$,
the original one.
While our starting model contained this state as a consequence of the interaction of the coupled channels,
it is unclear a priori, that the partial information of this interaction contained in the correlation functions is sufficient to determine the existence and position of that state.
Most remarkable is the fact that one can determine its mass with a precision of $6 \, \rm MeV$.
This contrast with the results of similar problems like the one of determining the position of the $D_{s0}^*(2317)$ state
from the $D^0K^+, D^+K^0$ and $D_s^+ \eta$ correlation functions,
where a binding energy of about $50\, \rm MeV$ is determined with a precision of $20\, \rm MeV$ \cite{Ikeno:2023ojl}.

There is more valuable information in Table \ref{tab:gi4}.
Indeed, the couplings are determined also with good precision and we see that $g_1/g_2=1.42$. This is remarkably close to $\sqrt{2}$,
which according to Eq.~\eqref{eq:iso} is what we could expect for a state being of $I=1/2$.
The fit to the correlation functions is, thus, telling us that the state obtained has $I=1/2$.
The uncertainties of the order of $10\%$ in the couplings cannot revert this fact since a state of $I=3/2$ would have this ratio negative and equal to $-1/\sqrt{2}$.

And now we face the problem of the probabilities of the state in each channel.
Before entering the discussion, let us note that the pole position is complex and it would correspond to a state that decays with a width of the order of $(178\pm 18)\, \rm MeV$, in line with the results of the PDG.
The width comes solely from the decay to the $\eta p$ channel which is open for the energy of the state.
The couplings are complex, but with imaginary parts smaller than the real parts.
The probabilities $\mathcal{P}_i$ calculated via Eq.~\eqref{eq:Pi} are complex and thus do not stand as probabilities.
The problem is more serious because the state, decaying to $\eta p$ will have an asymptotic wave function of the type $e^{ikr}/{r}$
which is not normalizable.
The meaning of this number is explained in Ref.~\cite{aceti}
and corresponds to the integral $\int d^3r \psi^2(r)$ (not $|\psi(r)|^2$) with a certain prescription for the phase of $\psi(r)$.
This is why the integral is finite since the integral of $e^{2ikr}/{r^2}$ cancels because of the oscillations.
Hence, one can interpret $\mathcal{P}_i$ as a strength of a channel,
given by the integral of the square of the wave function in the confinement region before the asymptotic one.
With this caveat one finds
\begin{equation}\label{eq:ppi}
  \begin{split}
   \mathcal{P}_1 \simeq & \; 0.12 -0.23 \,i, ~~~~~ \mathcal{P}_2 \simeq \, 0.06 -0.12 \,i, \\[1.5mm]
   \mathcal{P}_3 \simeq & \; 0.22 -0.28 \,i, ~~~~~ \mathcal{P}_4 \simeq  -0.34 -0.24 \,i, \\
  \end{split}
\end{equation}
and if one takes the modulus of these quantities we find
\begin{equation}\label{eq:pi2}
|\mathcal{P}_1| = 0.26, ~~~~ |\mathcal{P}_2| = 0.13, ~~~~ |\mathcal{P}_3| = 0.35, ~~~~ |\mathcal{P}_4| = 0.42.
\end{equation}
We find that the sum of these numbers, $1.16$, exceeds unity, which indicates again that these are not probabilities,
but gives us an idea or the strength of each channel.
We see that the two $K\Sigma$ channels together have a large strength,
in spite of being further away than the $K\Lambda$ from the mass of the state.
This further stresses the relevant role of the $K\Sigma$ channel as the backbone of the obtained state.
One can also see the nontrivial effect of coupled channels,
since the apparently innocuous $K\Lambda$ channel finally gets some component in the wave function through its coupling
$K\Lambda \to \eta p \to K\Sigma$.

It is also interesting to show the values of the wave function at the origin in coordinate space $g_i G_i$, given by
\begin{equation}\label{eq:psii}
  \begin{split}
   \psi_1(r=0) \, \simeq & \, -26+ 14 \, i, ~~~~~ \psi_2(r=0) \, \simeq \, -19+ 9.8 \, i, \\[1.5mm]
   \psi_3(r=0) \, \simeq & \, -30 + 11 \, i, ~~~~~ \psi_4(r=0) \, \simeq \, -18 -30 \, i. \\
  \end{split}
\end{equation}
The sum of the $K^0 \Sigma^+, K^+\Sigma^0$ components exceeds the $K^+\Lambda$ one,
stressing once more the importance of this channel.

In order to see the stability of the results, we have repeated the resampling method with $100$ runs.
We show the results for the scattering observables in Tables \ref{tab:ainew} and \ref{tab:rinew}, and for the couplings in Table \ref{tab:ginew}.
\begin{table}[t]
	\caption{Results with $100$ resampling runs for scattering lengths of channel $i$. [in units of fm]}
\centering
\begin{tabular*}{0.7\textwidth}{@{\extracolsep{\fill}}c| c }
\toprule
$a_{1}$         & $a_{2}$     \\[-1mm]
%\hline
~~~$(0.44\pm 0.05) -(0.62\pm 0.04) \, i$~~   & ~~$(0.31\pm0.02) -(0.34\pm 0.02) \, i$ ~~     \\%[1mm]
\hline\hline
  $a_{3}$  &  $a_{4}$ \\[-1mm]
%  \hline
   ~~~  $(0.30\pm0.02) -(0.20\pm 0.04) \,i$~~~   &  $-0.769\pm 0.017$ \\
\hline\hline
\end{tabular*}
\label{tab:ainew}
\end{table}
\begin{table}[t]
	\caption{Results with $100$ resampling runs for effective range parameters of channel $i$. [in units of fm]}
\centering
\begin{tabular*}{0.99\textwidth}{@{\extracolsep{\fill}}c| c| c| c }
\toprule
$r_1$         & $r_2$  &  $r_3$  &  $r_4$  \\
$(-1.2\pm 0.3) -(2.7\pm 0.2) \, i$   & $(-5.5\pm1.6)+ (8.9\pm 0.5) \, i$      & $(-2.8\pm 0.3) -(0.1\pm 0.7) \, i$   &  $-1.41\pm 0.16$  \\
\hline\hline
\end{tabular*}
\label{tab:rinew}
\end{table}
\begin{table}[bhtp]
     \renewcommand{\arraystretch}{0.9}
     \setlength{\tabcolsep}{0.3cm}
\centering
\caption{ Results with $100$ resampling runs for pole position and couplings. [in units of MeV]}
\label{tab:ginew}
\begin{tabular}{c|c|c}
\hline
\hline
$\sqrt{s_p}$ & $g_1$  & $g_2$  \\[-1mm]
   $(1515\pm 7) -(96\pm 13) \,i$  & $(3.7\pm0.5) -(1.11\pm 0.16) \, i$ & $(2.6\pm 0.4) -(0.79\pm 0.11) \, i$   \\
   \cline{2-3}
             & $g_3$ & $g_4$     \\[-1mm]
					   & $(3.5\pm 0.3) -(0.27\pm 0.06) \, i$ & $(-2.75\pm 0.17)+ (1.4\pm 0.2) \,i$    \\
\hline\hline
\end{tabular}
\end{table}

As we can see, the results are practically the same as those with $50$ runs, all compatible within the obtained uncertainties.
The results for the probabilities are also practically identical to those found in Eq.~\eqref{eq:pi2}.

\subsection{Comparison with experiment}
In Ref.~\cite{expelamb}, the $K^+\Lambda$ correlation function is analyzed in terms of the Lednick\'{y}-Lyuboshits analytical formula as a single channel \cite{Lednicky}.
A scattering length is obtained (in our convention $f(k)\simeq (-\frac{1}{a}+\frac{1}{2} r_0 k^2 -ik)^{-1}$)
\begin{equation}\label{eq:ai_exp}
  a=(0.61\pm 0.03 \pm 0.03) -i (0.23 \pm 0.06\pm 0.04) \; {\rm fm}.
\end{equation}
This should be compared with $a_3$ from Table \ref{tab:ai}
\begin{equation}\label{eq:ai_our}
  a^{\rm (ours)}=(0.30\pm 0.02)  -i (0.22 \pm 0.04) \; {\rm fm}.
\end{equation}
The imaginary parts are consistent, but the real part of $a$ is about twice our value of $a^{\rm (ours)}$. This apparent contradiction requires an explanation.
The real part of the scattering length is tightly linked to the value of the correlation function at $p=0$ in the Lednick\'{y}-Lyuboshits formula with a single channel
\begin{equation} \label{eq:CLL}
  C_{LL}(0)= 1+ \dfrac{2\, {\rm Re} f_0(0)}{\sqrt{\pi} \, R} + \dfrac{|f_0(0)|^2}{2\, R^2}.
\end{equation}
Using the source function of Ref.~\cite{expelamb}, one can see a fair agreement of Eq.~\eqref{eq:CLL} with the results of the correlation function of Ref.~\cite{expelamb} (Fig.~2 of that reference), $0.7$ versus $0.8$ (note also that they use Eq.~(2) to compare with the experimental data, slightly different than our theoretical formula).
Thus, the result for the scattering length obtained in Ref.~\cite{expelamb} with a single channel analysis is not surprising.
However, the correlation function for $K^+\Lambda$ that we show in Fig.~\ref{Fig:fig2} (left) is remarkable similar to the experimental one of Ref.~\cite{expelamb}.
Yet, we found that consistent with ${\rm Re}\,a^{\rm (ours)} \simeq 0.3 \, {\rm fm}$.
This apparent contradiction should be attributed to the effect of the coupled channels in Eq.~\eqref{eq:C3} which are ignored in the single channel analysis of Ref.~\cite{expelamb}.
We conclude then that the analysis of data within coupled channels is necessary and urge the experimental teams to measure the other correlation functions discussed in the present work.

\section{Conclusions}
 We address the problem of constructing the correlation functions of the $K^0 \Sigma^+, K^+ \Sigma^0, K^+ \Lambda$ and $\eta p$ channels.
 These channels, in particular the $K \Sigma$ channels, are responsible for the appearance of the $N^*(1535)$ state,
 that becomes dynamically generated by these channels within the chiral unitary approach.
 The interaction of these channels produces that state,
 but the $K^0 \Sigma^+, K^+ \Sigma^0, K^+ \Lambda$ channels are very far away in energy from the $N^*(1535)$
 and it is unclear how much the correlation functions of these channels,
 filtering information of the interaction above their respective thresholds,
 can determine the existence of a bound state  around $150 \, \rm MeV$ below the thresholds.
 We have then addressed the inverse problem,
 assuming that the correlation functions for the $K^0 \Sigma^+, K^+ \Sigma^0, K^+ \Lambda$ and $\eta p$ channels is known experimentally,
 for which we take synthetic data extracted from the evaluated correlation functions,
 and perform a fit to these data, very general and with minimal  assumptions.
 Then we obtain the values of the parameters of the framework used
 and with them we determine the different observables tied to the correlation functions.
 We determine the scattering length and effective range for each of the four channels
 and we find a pole in the $T$ matrix that corresponds to the $N^*(1535)$.
 The remarkable thing is not that we obtain this state,
 which was expected with the input used to obtain the correlation functions used in the fits,
 but that it is obtained with high precision.
 The uncertainty in the binding, obtained using the bootstrap method,
 is only $6 \, \rm MeV$.
 Very interesting also is that one can evaluate the couplings of the state to the different channels with also high precision,
 such that unmistakably one can assert that the state obtained has isospin $\frac{1}{2}$.
 The results obtained here should encourage the measurement of these correlation functions
 to see how much the $N^*(1535)$ is tied to the $K^0 \Sigma^+, K^+ \Sigma^0, K^+ \Lambda$ and $\eta p$ channels,
 a subject of much discussion concerning the nature of the state.
 For the moment, the $K^+\Lambda$ correlation function is available, which looks very similar to the one obtained here.
 However, when analyzed in the experiment by means of a single channel, the real part of the scattering length obtained has the same sign but is twice as big as the one obtained here, and we trace the discrepancy to the use of the single channel analysis in the experimental work \cite{expelamb}.
 We show the importance of using coupled channels in the analysis and make a call for the measurement of the complementary channels discussed in this work.

%\clearpage
\section{ACKNOWLEDGEMENT}
This work is partly supported by the National Natural Science Foundation of China under Grant No. 11975083 and No. 12365019, by the Natural Science Foundation of Guangxi province under Grant No. 2023JJA110076,
and by the Central Government Guidance Funds for Local Scientific and Technological Development, China (No. Guike ZY22096024).
This work is also supported partly by the Natural Science Foundation of Changsha under Grant No. kq2208257
and the Natural Science Foundation of Hunan province under Grant No. 2023JJ30647 (CWX).
R. M. acknowledges support from the CIDEGENT program with Ref. CIDEGENT/2019/015,
the Spanish Ministerio de Economia y Competitividad
and  European Union (NextGenerationEU/PRTR) by the grant with Ref. CNS2022-13614.
This work is also partly supported by the Spanish Ministerio de Economia y Competitividad (MINECO) and European FEDER
funds under Contracts No. FIS2017-84038-C2-1-P B, PID2020-112777GB-I00, and by Generalitat Valenciana under contract
PROMETEO/2020/023.
This project has received funding from the European Union Horizon 2020 research and innovation
programme under the program H2020-INFRAIA-2018-1, grant agreement No. 824093 of the STRONG-2020 project.

\end{document}